\documentclass[pra,twocolumn,amsmath,amssymb,floatfix,reprint,footinbib,superscriptaddress,longbibliography,showkeys]{revtex4-2}
\usepackage{picinpar,graphicx}
\usepackage{braket}
\usepackage{amsmath}
\usepackage{graphicx}
\usepackage{epstopdf}
\usepackage{dcolumn}
\usepackage{graphicx}
\usepackage{mathrsfs}
\usepackage{mdwlist}
\usepackage{subfigure}
\usepackage{booktabs}
\usepackage{amsmath}
\usepackage{dsfont}
\usepackage{amstext}
\usepackage{amssymb}
\usepackage{amsbsy}
\usepackage{bbm}
\usepackage{amsthm}
\usepackage{graphicx}
\usepackage{textcomp}
\usepackage{color}
\usepackage[colorlinks,citecolor=blue]{hyperref}
\usepackage{lipsum}
\definecolor{Dgreen}{RGB}{0, 100, 0}
\usepackage{url}
\usepackage[colorlinks]{hyperref}
\hypersetup{%
	plainpages=true,
	breaklinks=true,  
	hypertexnames=false,  
	pageanchor=true,
	colorlinks=true,
	linkcolor={blue},
	citecolor={red},
	urlcolor={blue},
	anchorcolor={black}
}

\begin{document}
	
	\title{Critical States Preparation With Deep Reinforcement Learning}
	\author{Jia-Wen Yu}
	\affiliation{Department of Physics, Fuzhou University, Fuzhou, 350116, China}
	\affiliation{Fujian Key Laboratory of Quantum Information and Quantum Optics, Fuzhou University, Fuzhou 350116, China}
	\author{Yi-Ming Yu}
	\affiliation{Department of Physics, Fuzhou University, Fuzhou, 350116, China}
    \affiliation{Fujian Key Laboratory of Quantum Information and Quantum Optics, Fuzhou University, Fuzhou 350116, China}
    \author{Ke-Xiong Yan}
	\affiliation{Department of Physics, Fuzhou University, Fuzhou, 350116, China}
    \affiliation{Fujian Key Laboratory of Quantum Information and Quantum Optics, Fuzhou University, Fuzhou 350116, China}
    \author{Jun-Hao Lin}
    \affiliation{Department of Physics, Fuzhou University, Fuzhou, 350116, China}
    \affiliation{Fujian Key Laboratory of Quantum Information and Quantum Optics, Fuzhou University, Fuzhou 350116, China}
    \author{Jie Song}
    \affiliation{Department of Physics, Harbin Institute of Technology, Harbin 150001, China}
    \author{Ye-Hong Chen}\thanks{yehong.chen@fzu.edu.cn}
	\affiliation{Department of Physics, Fuzhou University, Fuzhou, 350116, China}
    \affiliation{Fujian Key Laboratory of Quantum Information and Quantum Optics, Fuzhou University, Fuzhou 350116, China}
    \affiliation{Quantum Information Physics Theory Research Team, Center for Quantum Computing, RIKEN, Wako-shi, Saitama 351-0198, Japan}
    
    \author{Yan Xia}\thanks{xia-208@163.com}
	\affiliation{Department of Physics, Fuzhou University, Fuzhou, 350116, China}
    \affiliation{Fujian Key Laboratory of Quantum Information and Quantum Optics, Fuzhou University, Fuzhou 350116, China}

	\date{\today}
	
	\begin{abstract}
The fast and efficient preparation of quantum critical states is a challenging yet crucial task for various quantum technologies. This difficulty is most particularly for systems near a quantum phase transition, where the closure of the energy gap fundamentally limits the timescale of adiabatic processes and thus precludes rapid state preparation. We propose a framework using deep reinforcement learning (DRL) to rapidly prepare quantum critical states, with broad extendibility to light-matter interaction systems. Specifically, a DRL agent optimizes a set of time-dependent control Hamiltonians to drive the system from an initial noncritical state to a target critical state within a finite time and over experimentally accessible parameter ranges. As a concrete application, we focus on the quantum Rabi model. The DRL-optimized time-dependent control Hamiltonian yield a final state with high-fidelity ($>0.999$) to the target critical state. The protocol can be readily extended to other quantum critical systems described by light-matter interaction models, such as quantum Dicke model. This investigation provides a powerful new framework for preparing and manipulating quantum critical states.

	\end{abstract}
	
	\maketitle

	\textit{Introduction.}---Quantum systems near critical points exhibit a wealth of exotic physical phenomena, where small variations in physical parameters can lead to drastic changes in equilibrium-state properties~\cite{Scully_Zubairy_1997,Chen2024,ROSSINI20211,Sachdev_2011,science}. A prominent class of quantum critical systems is described by light-matter interaction models, such as the quantum Rabi model~\cite{PhysRevLett.115.180404,PhysRevA.97.013825}, the energy gap closes and levels converge to degeneracy as system parameters approach the critical point.
	
	 The ground states of these quantum critical systems are pivotal to the investigation of quantum phase transitions (QPT), as they exhibit quantum criticality-specific universal behavior, long-range correlations, and high entanglement~\cite{PhysRevLett.95.105701,PhysRevE.74.031123,CAROLLO20201}. These unique properties make them a vital resource for quantum technologies~\cite{Osterloh2002,PhysRevLett.90.227902,PhysRevLett.92.096402}, particularly in quantum metrology~\cite{PhysRevLett.124.120504,PhysRevLett.121.020402,PhysRevLett.130.240803}, where their exquisite sensitivity to parameter changes can dramatically enhance measurement precision. Nevertheless, preparing such critical ground states remains a formidable challenge. Conventional methods, such as adiabatic evolution~\cite{PhysRevA.87.063820,PhysRevB.89.094304}, demand infinitely slow passage through the critical point to avoid excitations induced by the closing energy gap, rendering them impractical for many applications. Consequently, fast, robust, and experimentally feasible protocols for generating quantum critical ground states is a crucial open challenge in the field~\cite{Innocenti_2020,PhysRevB.111.054311,https://doi.org/10.1002/qute.202100114}.
	
	The fast and high-fidelity preparation of a target quantum state can be formulated as an optimization problem~\cite{RevModPhys.58.1001,Handel_2005,PhysRevLett.122.010505,PhysRevA.99.063803,PhysRevA.106.033708}. Given a system's Hamiltonian, the challenge lies in engineering a time-dependent protocol for its control parameters to drive an initial state to a target state on timescales shorter than the system's decoherence time~\cite{Yu:25,PhysRevLett.126.023602,Mamaev2018dissipative,PhysRevLett.134.060601,PhysRevApplied.15.034068,Cardenas-Lopez2023,RevModPhys.91.045001}. Among existing solutions, efficient gradient-based optimal control methods (e.g., gradient-ascent pulse engineering~\cite{KHANEJA2005296,PhysRevA.84.022305,PRXQuantum.4.030305}) excel at identifying optimal control parameters via efficient gradient evaluation, but they rely on explicit knowledge of the system’s dynamics. However, in strongly or ultrastrongly coupled quantum critical systems (e.g., light-matter interaction models), the quantum dynamics are often analytically intractable and gradient calculations become unreliable—severe limitations that restrict gradient-based methods’ applicability. In such cases, conditioning the control sequence on final measurement outcomes (such as fidelity or physical observables) becomes essential~\cite{PRXQuantum.4.030305}.

	 Recently, deep reinforcement learning (DRL) has emerged as a powerful tool for tackling complex optimization tasks in quantum physics, particularly those involving measurement feedback~\cite{PhysRevLett.131.050601,PhysRevLett.126.060401,PhysRevLett.134.120803,PhysRevA.102.022412,PhysRevA.100.041801,PhysRevLett.131.073201,zhu2024controllingunknownquantumstates}. Deep reinforcement learning agents can learn effective control strategies through trial-and-error interaction with a simulated environment, without requiring prior knowledge of the system's intricate dynamics~\cite{8103164}. This approach has achieved remarkable success in various quantum domains, such as discovering protocols for preparing many-body ground states~\cite{PhysRevX.11.031070}, designing high-fidelity quantum gates~\cite{PhysRevA.110.032614}, and optimizing decoders for quantum error correction~\cite{PhysRevLett.133.150603}. Deep reinforcement learning excels at exploring complex, high-dimensional parameter landscapes to identify non-intuitive and highly efficient control protocols that are inaccessible to traditional analytical optimization.
	
	In this Letter, we propose a broadly applicable DRL-framework for the rapid preparation of quantum critical states: a DRL-agent optimizes  the time-dependent control Hamiltonians to drive a system from a non-critical ground state to a target critical state within a short finite time and high fidelity. As a concrete application, the quantum Rabi model (QRM) is considered, where the optimized control Hamiltonians generate a final state with a fidelity exceeding $0.999$ with respect to the target critical state using a single control field. To assess practical feasibility, the robustness of the control Hamiltonian sequence against systematic errors and environmental dissipation is examined. Random perturbations are applied to the parameters $\{\omega_d,\phi_i,\Lambda_i\}$ in Eq.~\eqref{control}, sampled from a standard normal distribution $\mathcal{N}(0,1)$. Numerical simulations show that such control errors reduce the state-preparation fidelity by less than $5\%$, while the influence of dissipation remains below $1\%$, confirming the robustness of the protocol.
	
	Furthermore, we validate the prepared state’s criticality via quantum Fisher information (QFI)~\cite{Liu_2020}: the QFI rises sharply toward the end of evolution, signifying the final state’s extreme sensitivity to system parameters—a hallmark of quantum criticality. This protocol establishes DRL as a powerful tool for engineering quantum critical states, with broad extendibility to other light-matter interaction systems (e.g., quantum Dicke model).

	\textit{Physical model and method.}---We consider a class of quantum systems whose Hamiltonian is given by
		\begin{align}
			H[g(t)]=H_{0}+H_{\rm{int}}[g(t)],
	\end{align}
	where $H_{0}$ denotes the system’s free Hamiltonian, $H_{\rm{int}}[g(t)]$ represents the interaction Hamiltonian of the quantum system, which depends on a tunable parameter $g(t)$~\cite{Innocenti_2020}. When $g(t)$ approaches the critical value $g_{\rm{c}}$, strong entanglement develops among the different degrees of freedom of the system, rendering controlled preparation of the critical state highly challenging. Given the initial and target values $g(0)$ and $g(T)=g_c$ over an evolution window $[0,T]$, the objective is to determine a time-dependent protocol that drives the system from the initial state $|\Phi[g(0)]\rangle$ to the eigenstate $|\Phi[g(T)]\rangle$ of the Hamiltonian $H[g(T)]$.
	
	The aforementioned state transition can be achieved via conventional methods: adiabatic evolution offers stability against experimental parameter fluctuations but requires infinitely slow passage through the critical point~\cite{PhysRevX.8.021022}; shortcuts to adiabaticity reduce the evolution time yet often rely on experimentally infeasible driving Hamiltonians~\cite{PhysRevA.93.052109} (see Supplemental Material for a comparison with our protocol~\cite{SM}). To address this, we introduce a sequence of control Hamiltonians. The total Hamiltonian then becomes
		\begin{align}
			H_{\rm tot}(t) = H[g(t)] + \sum_{i=1}^{n} H_i^{\rm c}(t),
		\end{align}
		where the control Hamiltonians take the form
		\begin{align}
			H_i^{\rm c}(t) = \Lambda_i \cos(\omega_d t + \phi_i)\, H_i^{\rm c}.
		\end{align}
		Here, $\omega_d$ denotes the driving frequency of the control fields, $\Lambda_i$ and $\phi_i$ are the corresponding amplitudes and phases, and $H_i^{\rm c}$ are the control Hamiltonian operators. In the first step, a set of control Hamiltonians $H_i^{\rm c}(t)$ is employed to provide full control over the degrees of freedom of the system and to realize quantum state transfer from $|\Phi[g(0)]\rangle$ to $|\Phi(T)\rangle$, while the system Hamiltonian is tuned from $H[g(0)]$ to $H[g(T)]$. In practice, however, it is challenging to implement all control fields simultaneously and address all degrees of freedom of the system. To overcome this limitation, we introduce the second step. Starting from the full set of optimized control parameters obtained in the first step, we apply each control field $i=1,2,\ldots,n$ individually to drive the system dynamics and obtain a set of test states $|\Phi^{\mathrm{test}}_{i}(t)\rangle$. We then define a trajectory similarity
		\begin{align}
			\Delta_{i} = \int_{0}^{T} \big|\langle \Phi^{\mathrm{test}}_{i}(t) | \Phi(t) \rangle \big|^{2} \, dt ,
			\label{similarity}
		\end{align}
		where $\Delta_{i}$ quantifies the contribution of the $i$-th control field to the target evolution. Based on the values of $\Delta_{i}$, control fields with negligible contributions are discarded. The remaining dominant control fields are subsequently reoptimized to further reduce resource consumption while preserving the target state transfer.
		
	\begin{figure}
		\centering
		\includegraphics[scale=0.46]{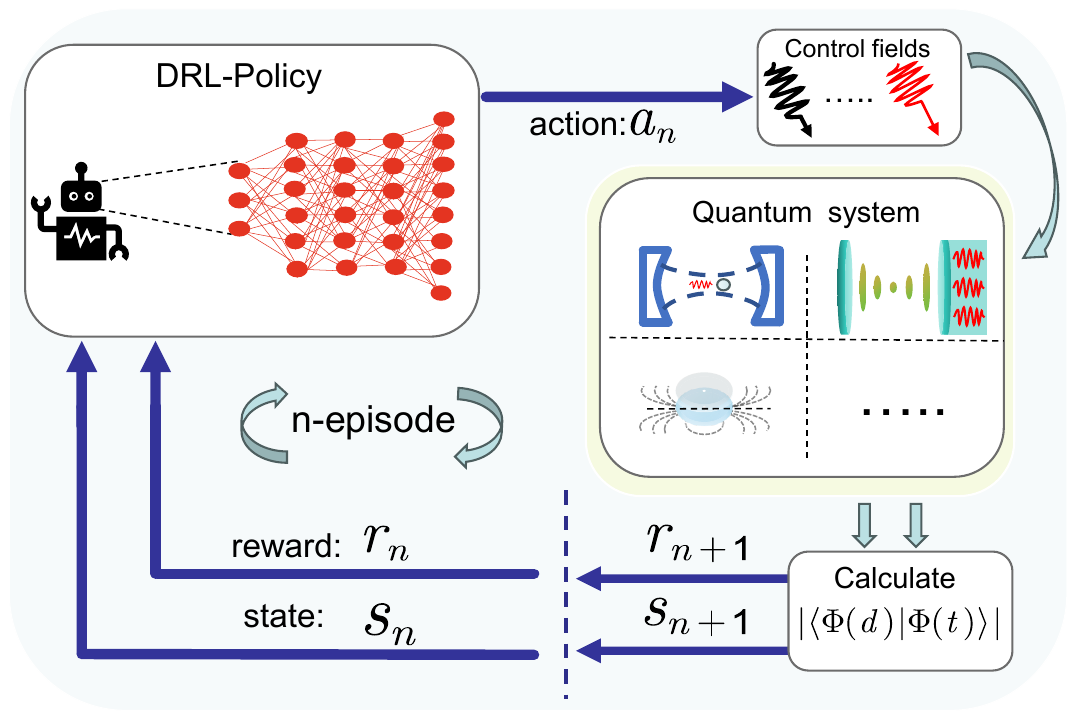}
		\caption{Schematic of the DRL approach for quantum critical state preparation. 
			At each episode $n$, the DRL agent obtains the current state $s_{n}$ of the quantum system and the reward $r_{n+1}$, 
			defined by the fidelity between the driven final state $|\Phi(d)\rangle$ and the target state $|\Phi(t)\rangle$ 
			(such as photon--atom, photon--phonon, or magnon systems). 
			Using its policy function $\pi(A|S)$, the agent selects an action $a_{n}$ that specifies a configuration of time-dependent control fields. 
			The selected control is then applied to the quantum system, driving its evolution. 
			This control is applied to the system, driving its evolution to a new state $s_{n+1}$ and yielding a new reward $r_{n+1}$. 
			This iterative loop enables the DRL agent to optimize the control Hamiltonians, guiding the system to the target critical state.
		}
		\label{fig1}
	\end{figure}
	
	In general, determining a sequence of control Hamiltonians $H_{i}^{c}(t)$ that drives a complex quantum system to a desired evolution is challenging. The system dynamics are rarely analytically solvable as the $g(t)$ increases. Moreover, in the first step full control analysis, one must explore a high dimensional parameter space, which leads to a substantial optimization overhead. We therefore employ DRL to design control protocols that circumvent these analytical limitations. The central idea is that the DRL model (workflow in Fig.\ref{fig1}) emulates the behavior of agents interacting with an environment, learning through rewards to adapt its decisions based on observations~\cite{PhysRevA.110.032614}. 
	
	In the learning process, a fixed total evolution time $T$ is discretized into a finite number of steps $k=1,2,...,K$. At each episode $n$, the agent observes the current state $s_{n}\in S$, and choose an action $a_{n}\in A$ according to the policy $\pi(A|S)$. The action spaces are defined by the control parameters of the control Hamiltonians $H_{{i}}^{c}(t)$, such as driving frequency, amplitudes,  driving time and phases, and the state $s_{n}$ is described by the fidelity $F=|\langle\Phi(T)|\Phi(g_{t})\rangle|$ (Note that we adopt this form of the fidelity to mitigate the barren plateau problem that may arise during the initial training stage due to low-fidelity values). Other distance measures, such as the Jensen-Shannon divergence and the trace distance, can also be used to characterize the state $s_{n}$. However, near the critical point these measures require significantly larger computational resources. Therefore, the fidelity is chosen here as the training objective. To find the optimal the control Hamiltonians $H_{{i}}^{c}(t)$, we define the reward function for the DRL agent as 
		\begin{align}
			R&=r_{\rm{fid}}-\zeta_{\rm{amp}}P_{\rm{amp}}-\zeta_{\rm{freq}}P_{\rm{freq}}-\zeta_{\rm{smooth}}P_{\rm{smooth}},\nonumber\\
			r_{\rm{fid}}&=[a+b(F^{4})][-\log_{10}(1-F)],\nonumber\\
			P_{\rm{amp}}&=\sum_{i=1}^{n}[\frac{1}{n}\sum_{k=1}^{K}\exp(\varLambda_{k,i})],\nonumber\\
			P_{\rm{freq}}&=\exp(\omega_{d}),\nonumber\\
			P_{\rm{smooth}}&=\sum_{i=1}^{n}[\sum_{k=1}^{K}(\varLambda_{k+1,i}-\varLambda_{k,i})^2+2\sum_{k=1}^{K}(\varLambda_{k+2,i}\nonumber\\
			&-2\varLambda_{k+1,i}+\varLambda_{k,i})^2].\nonumber\\	
		\end{align}
		Here, the term $r_{\rm fid}$ denotes the fidelity reward and constitutes the primary driving component of the reward function $R$, $a$ and $b$ are positive hyperparameters (chosen to balance training stability and reward sensitivity). The penalty term $P_{\rm amp}$ suppresses excessive driving amplitudes, $P_{\rm freq}$ biases the agent toward lower driving frequencies, and $P_{\rm smooth}$ reduces abrupt changes in the driving amplitude between adjacent time steps. The coefficients $\zeta_{\rm amp}$, $\zeta_{\rm freq}$ and $\zeta_{\rm smooth}$ represent the weights of the corresponding penalty terms. The structure of the reward function enforces strong physical constraints during training.
	
	We train the agent using the proximal policy optimization (PPO) algorithm~\cite{schulman2017proximalpolicyoptimizationalgorithms} 
	to maximize the cumulative reward over the control sequence. The implementation is based on the open-source \texttt{TensorForce} library. A specific example is presented below.

	\textit{Application-Quantum Rabi model.}---To validate our DRL framework, we focus on the quantum Rabi model (QRM)—a paradigmatic light-matter interaction system exhibiting a superradiant quantum phase transition~\cite{PhysRevLett.115.180404,PhysRevA.97.013825}. Its Hamiltonian is
		\begin{align}
			H_{\rm{Rabi}}[g(t)]=\omega a^{\dagger}a+\frac{\Omega}{2}\sigma_{z}+\frac{g(t)\sqrt{\omega\Omega}}{2}(a+a^{\dagger})\sigma_{x},
	\end{align}
	where $\sigma_{x,z}$ are Pauli matrices for the two-level atom, $\omega$ is the cavity field frequency, $\Omega$ is the energy splitting of a two-level atom,  and $a^{\dagger}(a)$ is the creation (annihilation) operator. We define the dimensionless parameter $g(t)=2\lambda(t)/\sqrt{\omega\Omega}$, where $\lambda(t)$ denotes the time-dependent cavity-atom coupling strength. The function $g(t)$ is parameterized as $g(t)=\sum_{j=0}^{2} a_j t^j$ and satisfies the boundary conditions $g(0)=g_0$ and $g(T)=g_{\rm{c}}$. In the limit of $\omega/\Omega \to 0$ thermodynamic, the ground state of $H_{\mathrm{Rabi}}(g)$ for $g\leq1$ is $|\Phi(g)\rangle = S[r(g)]|0\rangle |\!\downarrow\rangle$, with $S[r] = \exp\left[{r}/{2}(a^{\dagger 2} - a^2)\right]$ and $r(g) = -\frac{1}{4}\ln(1-g^2)$~\cite{PhysRevLett.115.180404}. The energy gap above the ground state close and the QRM exhibits a superradiant phase transition at $g_{\rm{c}}=1$, i.e., $\lambda=\sqrt{\omega\Omega}/2$.

    Here, five commonly used control Hamiltonians in cavity quantum electrodynamics systems are selected, providing full control over the degrees of freedom of the QRM ~\cite{Walther_2006,doi:10.1126/science.1078446}:
	\begin{align}
		H_{{1}}^{c}(t)&=\varLambda_{1}\cos{(\omega_{d}t+\phi_{1})}(a+a^{\dagger}),
		\nonumber\\H_{{2}}^{c}(t)&=\varLambda_{2}\cos{(\omega_{d}t+\phi_{2})}(a+a^{\dagger})^2,\nonumber\\H_{{3}}^{c}(t)&=\varLambda_{3}\cos{(\omega_{d}t+\phi_{3})}a^{\dagger}a,\nonumber\\H_{{4}}^{c}(t)&=\varLambda_{4}\cos{(\omega_{d}t+\phi_{4})}\sigma_{z},\nonumber\\H_{{5}}^{c}(t)&=\varLambda_{5}\cos{(\omega_{d}t+\phi_{5})}\sigma_{x},
		\label{control}
	\end{align}
	where we set $\omega_{d}$ and $\phi_{i}$ to be constant throughout the evolution process. A fixed total evolution time $T$ is discretized into a finite number of $k$ steps, and in each episode, we determine a set of amplitude sequences $\varLambda_{i}=\{\varLambda_{i}(t_{1}),\varLambda_{i}(t_{2}),..., \varLambda_{i}(t_{K})\}$. Considering realistic experimental constraints, we impose the conditions $\varLambda_{i}(t_{1})=\varLambda_{i}(t_{K})=0$ to ensure that $H_{{i}}^{c}(t)$ vanishes at the beginning $(t=0)$ and end $(t=T)$ of the evolution.
	
	At each DRL episode, the agent observes the state from the previous episode and selects a new set of parameters $\{T,\omega_d,\phi_i,\Lambda_i\}$ from the action space, which is preprocessed to constrain experimentally feasible parameters ranges. Consequently, the evolution operator can be described as
		\begin{align}
			U(T,0)=\prod_{k=K}^{1}U(t_{k},t_{k-1}),
		\end{align}
		where the single-step evolution operator takes the form
		\begin{align}
			U(t_{k}, t_{k-1}) = &\mathcal{T} \exp \biggl[-i \int_{t_{k-1}}^{t_{k}} \Bigl( H_{\mathrm{Rabi}}[g(\tau)] \\
			&+ \sum_{i=1}^{5} H_{i}^{c}(\tau) \Bigr) d\tau \biggr].
		\end{align}
		Here, $\mathcal{T}$ denotes the time-ordering operator. The final state of the quantum system is given by $|\Phi(T)\rangle = U(T,0)|\Phi[g(0)]\rangle$. Through repeated training, the agent adjusts the system evolution operator to maximize the fidelity $F(|\Phi(T)\rangle,|\Phi(g_{\rm{c}})\rangle)$. The trajectory similarity defined in Eq.~\eqref{similarity} is then analyzed to identify the control field with the largest contribution, which is subsequently selected for retraining in order to maximize the fidelity while minimizing resource consumption. Systemic state and reward are obtained by simulating the Schrödinger equation using the \texttt{QuTiP} package~\cite{JOHANSSON20121760,JOHANSSON20131234}. The collected states and rewards during the sampling process are used to update the policy via the PPO algorithm.
		
			\begin{figure}
			\centering
			\includegraphics[scale=0.47]{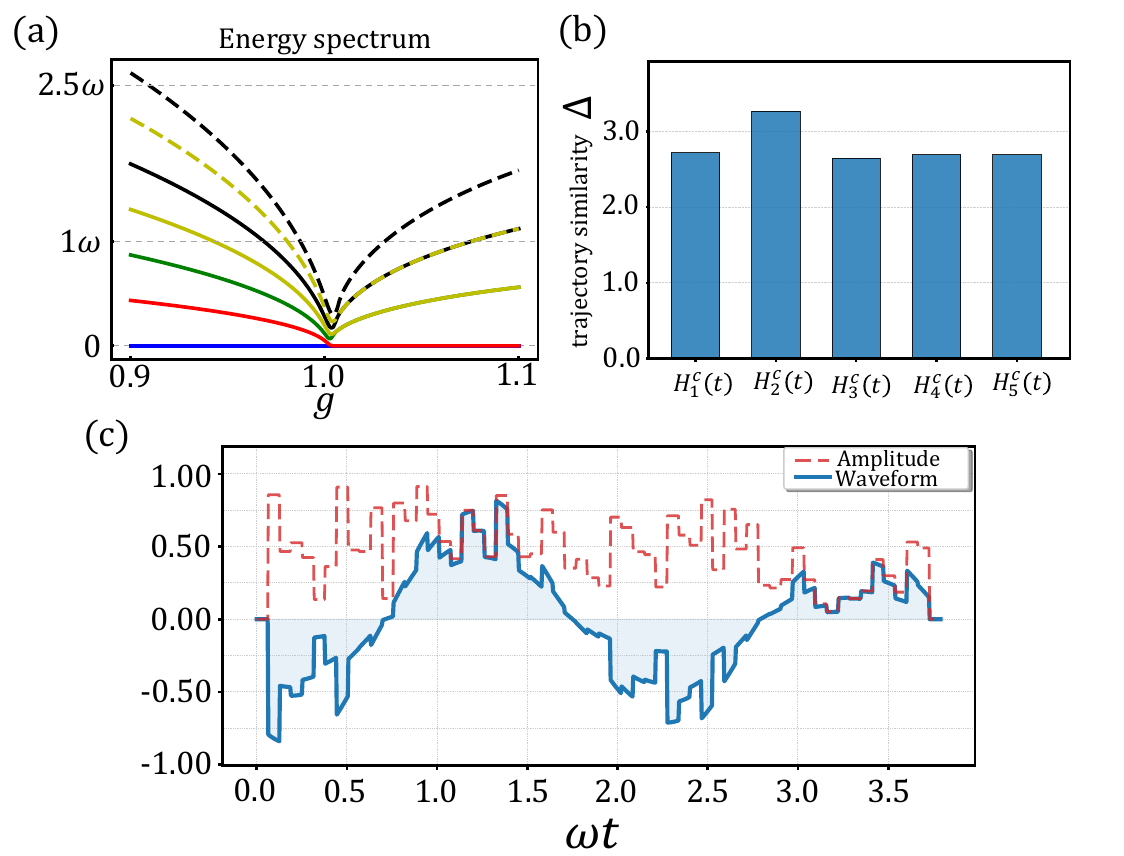}
			\caption{(a) Energy spectrum $(E_n - E_0)$ of the Rabi Hamiltonian $H_{\rm Rabi}[g(t)]$ across the critical point, calculated with a Hilbert-space truncation dimension of $N = 5000$.
					(b) Analysis of the trajectory similarity $\Delta_i$, indicating that the control field $(a + a^{\dagger})^2$ plays a dominant role in the system evolution.
					(c) Waveform and amplitude of the control pulses identified by the DRL-agent. The control Hamiltonian is given by $H^{\rm{c}}_{2}(t) = f_2(t)(a + a^{\dagger})^2$, with the driving field $f_2(t) = \varLambda_{2} \cos(\omega_d t + \phi_2)$. The parameters used in the simulation are $\omega = 1$, $\Omega = 10^4\omega$, $g_0 = 0.01$, $g_{\rm{c}} = 1$, and $K = 60$.}
			\label{fig2}
		\end{figure}
	
	\begin{figure}
		\centering
		\includegraphics[scale=0.76]{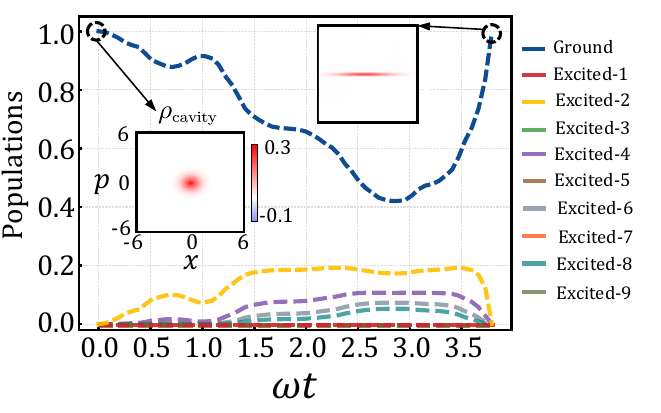}
		\caption{Time evolution of the populations. 
				At each time $t$, the evolving state $|\Phi(t)\rangle$ is projected onto the first ten eigenstates $|E_{n}(t)\rangle$ 
				of the instantaneous Hamiltonian $H_{\rm{Rabi}}[g(t)]$, yielding the population distribution 
				$P_{n}(t)=|\langle E_{n}(t)|\Phi(t)\rangle|^2$. 
				We also show the Wigner functions of the initial and final states of the cavity subspace, 
				obtained by tracing out the qubit degrees of freedom.
		}
		\label{fig6}
	\end{figure}
	
	In the first step, the DRL–optimized control sequence with the full set of control fields achieves a fidelity of $F(|\Phi(T)\rangle,|\Phi(g_{\rm{c}})\rangle)=0.9990$ at $\omega T = 4.1531$. The trajectory similarity $\Delta_2$ attains the largest value, indicating that the control field $(a + a^{\dagger})^2$ plays a dominant role in the evolution. Based on this result, the second step employs only the control field $(a + a^{\dagger})^2$ for the evolutionary training. The resulting optimized control sequence achieves the same fidelity, $F(|\Phi(T)\rangle,|\Phi(g_{c})\rangle)=0.9991$, at a shorter duration of $\omega T = 3.7913$.
	
	The corresponding optimal parameters $\{\omega_{d},\phi_{2},\varLambda_{2}\}$ are shown in Fig.~\ref{fig2} (In the Supplemental Material~\cite{SM}, control sequences for different system parameters $\Omega / \omega$ and the smoothing of the control waveforms are presented). 
	The populations of the time-evolved state $|\Phi(T)\rangle$ in terms of the instantaneous eigenstates of $H_{\rm{Rabi}}[g(t)]$
	are presented in Fig.~\ref{fig6}. The results show that DRL-agent identifies a shortcut distinct from the adiabatic path, enabling the system to evolve to the critical ground state $|\Phi(g_{\rm{c}})\rangle$ within a markedly short time. This evolution is achieved using a single driving field $(a + a^{\dagger})^2$, with the amplitude remaining within experimentally accessible limits and the duration of each single-stage operation $t_{k}$ well within the range of conventional quantum control times (In the Supplemental Material~\cite{SM}, a complete scheme for a potential experimental implementation is presented, which demonstrates the feasibility of the proposed protocol).
	
 	\textit{Analysis of noises.}---Without loss of generality, the influence of control errors and environmental dissipation on the DRL-control sequence is considered separately.
 	
\begin{figure}
 		\centering
 		\includegraphics[scale=0.6]{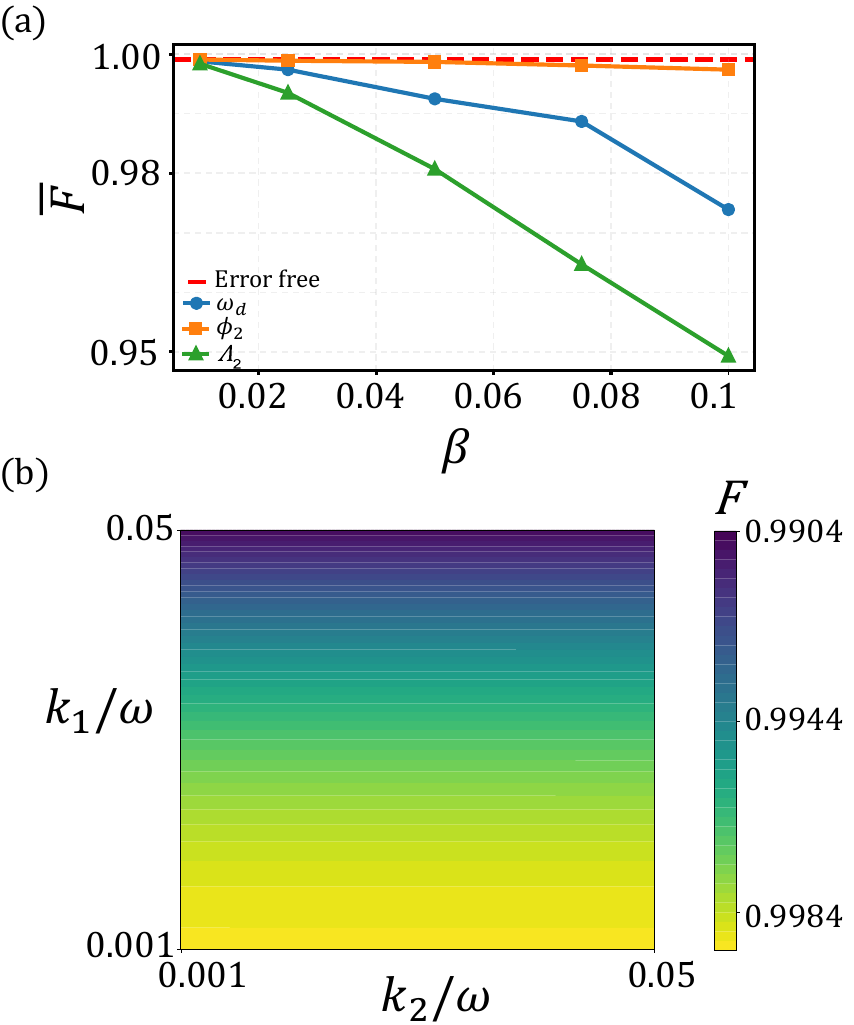}
 		\caption{(a) Analysis of control errors. Different broken lines correspond to errors in different control parameters $\chi$. The results show that the maximum fidelity loss of the final evolved state due to systematic errors remains below $5\%$. (b) Analysis of dissipative effects. The dephasing rate of the qubit is fixed at $\kappa_3/\omega = 0.01$. The results indicate that qubit relaxation has a negligible impact on the final evolved state, and even at relatively large dissipation rates, the fidelity between the final state and the target state remains above $0.99$.}
 		\label{fig7}
 	\end{figure}

 	For systematic errors, the actual control Hamiltonian is written as
 	\begin{align}
 		H_2^{\mathrm{ac}}(t)=H_2^{\mathrm{c}}(t)+H_2^{\mathrm{e}},
 	\end{align}
 	where $H_2^{\mathrm{e}}$ denotes the control error Hamiltonian. In practice, imperfections may occur in one or more control parameters $\{\omega_d,\phi_2,\varLambda_{2}\}$. To assess the robustness of the protocol, we examine the influence of deviation in each parameter independently while keeping the others fixed at their optimal values. A perturbed parameter is modeled as
 	$\chi^{a}=\chi+\beta\varepsilon$,
 	where $\chi\in\{\omega_d,\phi_2,\varLambda_{2}\}$, $\beta$ characterizes the relative error strength and $\varepsilon$ is a random variable drawn from a normal distribution $\mathcal{N}(0,1)$. The reported fidelity $\overline{F}$ is obtained by averaging over 100 independent realizations to suppress statistical fluctuations. In Fig.~\ref{fig7}(a), the effects of systematic errors in different parameters are displayed. The results show that the maximum fidelity loss of the final evolved state due to systematic errors remains below $5\%$, indicating that the control sequence exhibits high tolerance to system errors.
 	 
 	We further account for environmental dissipation during the system evolution. The dynamics is described by a Lindblad master equation~\cite{PhysRevLett.124.120504}
 	\begin{align}
 		\frac{d\rho(t)}{dt}
 		=&-i[H_{\rm{tot}}(t),\rho(t)] \nonumber\\
 		&+\sum_{i=1}^{3}\left(
 		L_i\rho(t)L_i^{\dagger}
 		-\frac{1}{2}\{L_i^{\dagger}L_i,\rho(t)\}
 		\right),
 	\end{align}
 	where $H_{\rm{tot}}(t)=H_{\rm{Rabi}}[g(t)]+H_2^{\mathrm{c}}(t)$ and $\rho(t)$ is the system density matrix. The Lindblad operators
 	$L_1=\kappa_1 a$, $L_2=\kappa_2\sigma_-$, and $L_3=\kappa_3\sigma_z$
 	correspond to photon loss, qubit relaxation, and qubit dephasing, respectively, with $\kappa_1$, $\kappa_2$, and $\kappa_3$ the associated decay rates. The effects of environmental dissipation are summarized in Fig.~\ref{fig7}(b). The results indicate that the protocol maintains high fidelity ($>0.99$) even in the presence of strong dissipation. In addition, we further apply the DRL-framework to train the control sequence directly in an open quantum system. In this case, the system state and reward are obtained by numerically simulating the Lindblad master equation. The highest fidelity between the final evolved state and the target state achieved by the trained control sequence reaches $F = 0.9965, \omega T = 2.6047$ (the parameters used in the simulation are $\omega = 1$, $\Omega = 100\omega$, $g_0 = 0.01$, $g_{\mathrm{c}} = 1$, $K = 60$, and decay rates $\kappa_1/\omega = \kappa_2/\omega = \kappa_3/\omega = 0.05$). These results demonstrate that the physically constrained reward function and the DRL environment designed in our work remain effective even in the presence of dissipation.

 	These features demonstrate the robustness of the designed control scheme and its suitability for realistic experimental implementations.

	\textit{Quantum Fisher Information.}---Quantum metrology relies on quantum Cramér-Rao bound, which limits the sensitivity of estimating an unknown parameter $\Theta$~\cite{Gietka2021adiabaticcritical}
	\begin{align}
		\rm{Var[\Theta]}\geqslant\frac{1}{\sqrt{{\textit{I}}_{\Theta}}}
	\end{align}
	where the QFI ${{I}_{\Theta}}$ for a pure state $|\Phi\rangle$ with respect to a parameter $\Theta$ is
	\begin{align}
		{I}_{\Theta}=4(\langle\partial_{\Theta}\Phi|\partial_{\Theta}\Phi\rangle-|\langle\partial_{\Theta}\Phi|\Phi\rangle|^{2}),
		\label{QFI}
	\end{align}
	with $\partial_{\Theta}=\partial/\partial\Theta$. 
	
	\begin{figure}
		\centering
		\includegraphics[scale=0.42]{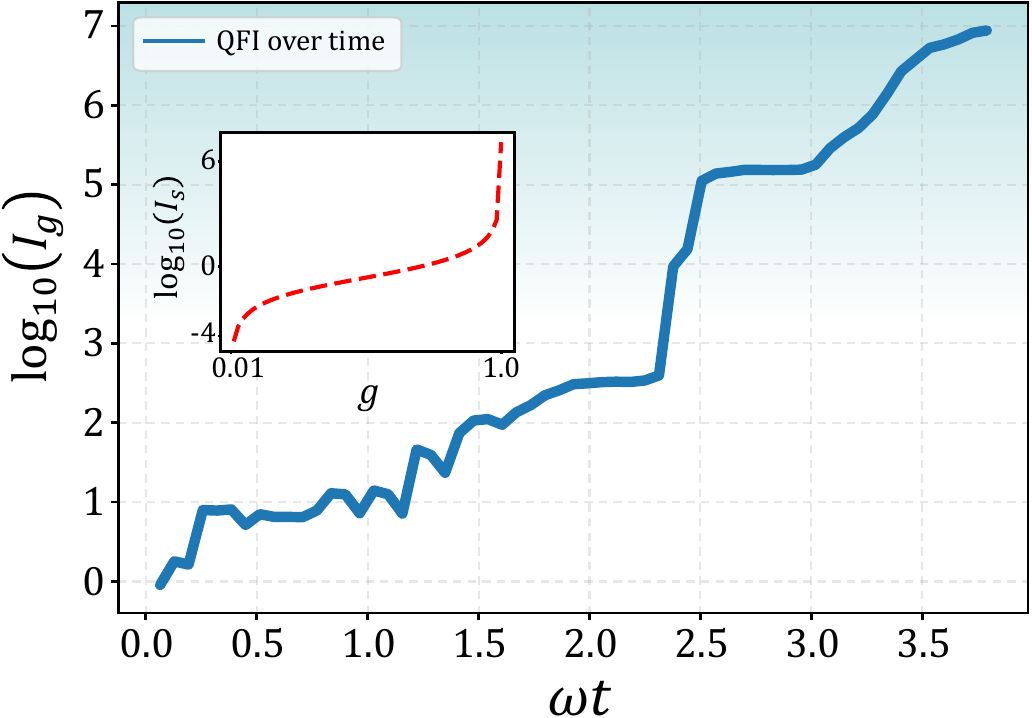}
		\caption{Time evolution of $I_g$ calculated for the evolving state $|\Phi(t)\rangle$ under the DRL-optimized control protocol, 
				together with $I_s={g^2}/{2(1-g^2)^2}$ in the limit $\omega/\Omega \to 0$, shown as the standard upper bound. 
				The results demonstrate that as the evolution approaches the final time ($t \to T$), the final state acquires extreme sensitivity 
				to the parameter $g$ and $I_g$ approaches $I_s$. 
				Here $\delta = 10^{-5}$ and $g_{s}=0.9999$.
		}
		\label{fig4}
	\end{figure}
		To verify that the prepared state exhibits criticality-specific parameter sensitivity, a hallmark of quantum critical states, the quantum Fisher information with respect to the parameter $g(t)$ is analyzed. In the following discussion, the time dependence of $g$ is suppressed for clarity. The states $|\Phi(t)\rangle=|\Phi(t,g)\rangle$ during evolution are implicitly $g$-dependent, so we approximate the derivative via the finite difference method~\cite{Zhou1993}. Therefore, we set the estimating parameter $\Theta=g$ and calculate the QFI of the states $|\Phi(t,g)\rangle$ at any time
	\begin{align}
		\frac{\partial|\Phi(t,g)\rangle}{\partial g}\thickapprox \frac{|\Phi(t,g+\delta)\rangle-|\Phi(t,g)\rangle}{\delta},
		\label{app}
	\end{align}
	where $\delta$ is a small perturbation. The physical meaning of Eq. $\eqref{app}$ is to quantify how much the geometric shape of the quantum state $\Phi(t,g)\rangle$ (its position in Hilbert space) "shifts" or "distorts" at a given time $t$ due to a minute change in the parameter $g$. Consequently, we obtain the complete QFI ${I}_{g}$ and its time-dependent trajectory by performing this calculation at each time point $t$.
	
	Figure \ref{fig4} presents the results of our numerical simulation. The value of curve ${I}_{g}$ sharply increases towards divergence at the end of the evolution, implying that the final state has acquired the heightened sensitivity to parameter variations that is characteristic of a critical point.

	\textit{Discussions and conclusions.}---We propose a deep DRL-based framework for fast, robust preparation of quantum critical states. The core DRL agent to optimize time-dependent control Hamiltonians $H_{{i}}^{c}(t)$ to driving a quantum system from an initial non-critical ground state $|\Phi[g(0)]\rangle$ to the target critical state $|\Phi[g(T)]\rangle$ ($g(T)=g_{\rm{c}}$) within a specified time. As an application, we investigated the QRM which exhibits a second order mean-field QPT at a critical value $g_{c}=1$. Optimal parameters found by the DRL algorithm yield a fidelity of $F(|\Phi(T)\rangle,|\Phi(g_{c})\rangle)=0.9991$ at $\omega T=3.7913$. To address more realistic scenarios, the robustness of the obtained control Hamiltonian sequence against systematic errors and environmental dissipation is analyzed. Random errors are introduced in control parameters $\{\omega_d,\phi_2,\Lambda_2\}$, drawn from a standard normal distribution $\mathcal{N}(0,1)$. Numerical simulations show that the state-preparation fidelity is reduced by less than $5\%$ compared with the error-free case, while the influence of dissipation remains below $1\%$, confirming the robustness of the protocol against external noise. QFI analysis further confirms the final state’s criticality: ${I}_{g}$ sharply increases towards divergence at the end of the evolution, reflecting the critical phenomenon of the QPT point. Importantly, the proposed framework is readily extendible to other light–matter quantum critical systems, such as the quantum Dicke model. For the quantum Dicke model, the protocol yields a fidelity $F(|\Phi(T)\rangle,|\Phi(g_c')\rangle)=0.9953$ at $\omega T = 3.0279$, and the corresponding optimal parameters are shown in Fig.~\ref{fig8}. Additional details are provided in the Supplemental Material~\cite{SM}, highlighting the broad applicability of the framework beyond the QRM. These findings collectively demonstrate the DRL's effectiveness and potential in precisely preparing and manipulating quantum critical states. 
	 \begin{figure}
		\centering
		\includegraphics[scale=0.54]{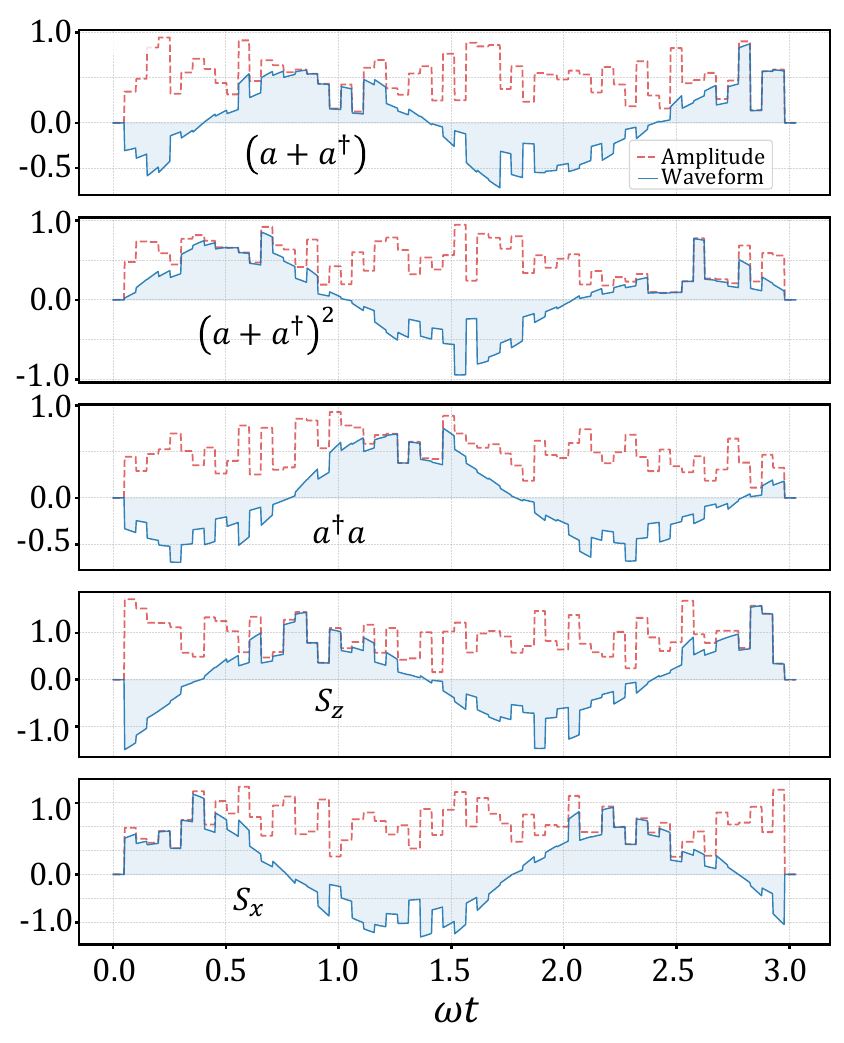}
		\caption{Waveforms of the control pulses discovered by the DRL agent. The parameters used in the simulation are $\omega^{'} = 1$, $\Omega^{'} = 2500\omega$, $N=4$, $g^{'}_0 = 0.01$, $g^{'}_{\rm{c}} = 1$, and $K = 60$.}
		\label{fig8}
	\end{figure}
	Notably, our framework leverages DRL in a simulated environment, it does not constitute a fully model-free approach~\cite{PhysRevX.12.011059}. For quantum control near critical points, a strictly model-free RL protocol would require experimental access to fidelity-based rewards and high-dimensional state information, which is extremely challenging and costly. Nevertheless, our results provide a foundation for future directions toward hybrid or hardware-efficient reward definitions that may enable partially model-free training in experiments.
	
\section*{{Acknowledgements}}
Y.-H.C. was supported by the National Natural Science Foundation of China under Grant No. 12304390 and 12574386, the Fujian 100 Talents Program, and the Fujian Minjiang Scholar Program. 
Y.X. is supported by the National Natural Science Foundation of China
under Grant No. 11575045 and No. 62471143, the Natural Science Funds
for Distinguished Young Scholar of Fujian Province under
Grant 2020J06011 and Project from Fuzhou University
under Grant JG202001-2.

\section*{Author contributions}
Y.-H.C. conceived and developed the idea. J.-W.Y., K.-X.Y., Y.-M.Y., and J.-H.L. analyzed the data and
performed the numerical simulations, with help from Y.X., J.S., and Y.-H.C.. J.-W.Y. and Y.-H.C. cowrote the paper with feedback from all authors.

\section*{Data availability}

The data used for obtaining the presented numerical results as well as for generating
the plots is available on request. Please refer to yehong.chen@fzu.edu.cn

\section*{Competing interests}

The authors declare that they have no competing interests.

\textit{}
\bibliography{reference}

\end{document}